\documentclass{aa}  
\usepackage{graphicx}
\usepackage{txfonts}
\usepackage{lipsum}
\usepackage{subcaption}        
\usepackage[version=4]{mhchem}
\usepackage{lscape}
\usepackage{placeins}
\usepackage{natbib}
\bibpunct{(}{)}{;}{a}{}{,}
\usepackage{booktabs}
\usepackage{threeparttable}
\usepackage{chemformula}
\usepackage{siunitx}
\usepackage[colorlinks=true, linkcolor=blue, citecolor=blue, urlcolor=blue]{hyperref}
\usepackage{stfloats}
\usepackage{textcomp} 
\usepackage{makecell}

\providecommand{\apj}{ApJ}
\providecommand{\apjl}{ApJL}
\providecommand{\apjs}{ApJS}
\providecommand{\aap}{A\&A}

\begin{document}

\titlerunning{ML Diagnostics for Interstellar PAHs}
\authorrunning{Wang}

\title{Full-spectrum infrared fingerprinting: A transformative AI paradigm for interstellar polycyclic aromatic hydrocarbons}

\author{Zhao Wang \inst{1}\fnmsep\thanks{Corresponding author: \email{zw@gxu.edu.cn}}}

\institute{Laboratory for Relativistic Astrophysics, Department of Physics, Guangxi University, 530004 Nanning, China}

\date{ARXIV}

\abstract
{In the era of high-sensitivity infrared (IR) astronomy, traditional manual diagnostics are no longer sufficient to harvest the complex physical insights hidden within interstellar spectra.}
{We introduce a machine learning paradigm that bypasses the limitations of empirical band ratios by treating the complete IR spectrum of polycyclic aromatic hydrocarbons (PAHs) as a high-dimensional fingerprint.}
{Using a random forest classifier trained on $\sim$\num{23000} spectra, we achieved a robust $F_{1}$ score of $\sim$0.963 across 12 size and charge categories, maintaining high performance on unseen molecular mixtures.}
{Interrogating the model's decision-making process reveals that PAH size diagnostics are charge-dependent. Neutral PAHs are traced by \ce{C-H} modes, while ionized species rely on 6--\SI{8}{\micro\meter} \ce{C-C} morphology; however, the \SI{12.5}{\micro\meter} feature remains a versatile tracer across multiple charge states.}
{This AI-driven paradigm offers a new route to interpret IR signatures and probe the chemical complexity of the interstellar medium.}

\maketitle
 \nolinenumbers
\section{Introduction}

Interstellar polycyclic aromatic hydrocarbons (PAHs) are primary subjects in astrophysical exploration; they are studied via their infrared (IR) emission \citep{Leger1984, Allamandola1985}. Their characteristic mid-IR (MIR) features (notably at 3.3, 6.2, 7.7, 8.6, 11.2, and \SI{12.7}{\micro\meter}) are ubiquitous, appearing in diverse environments ranging from individual stellar sources to entire galaxies \citep{Peeters2002, Smith2007}. Since PAH sizes and ionization states are sensitive to the local ultraviolet field and electron density, these molecules serve as vital diagnostics for energetic regions such as photodissociation regions and active galactic nuclei \citep{Tielens2008, Li2020}. 

Historically, inferring PAH properties relied on empirical band ratios calibrated against limited laboratory or theoretical datasets \citep{Allamandola1989, Draine2007}. These ratios are rooted in vibrational physics, and  the $I_{11.2}/I_{3.3}$ ratio serves as a primary proxy for PAH size. This is because smaller grains reach higher peak temperatures during stochastic heating, preferentially exciting the \SI{3.3}{\micro\meter} \ce{C-H} stretch over the \SI{11.2}{\micro\meter} mode \citep{Draine2001}. Similarly, ratios such as $I_{6.2}/I_{11.2}$ (or $I_{7.7}/I_{11.2}$) and $I_{11.2}/I_{12.7}$ are standard tools for tracing ionization states and molecular edge structures, respectively \citep{Hony2001, Galliano2008, Boersma2018}. Despite the evolution from the ``blind'' mathematical decomposition to a template-based fitting approach using the NASA Ames PAH IR spectroscopic database (PAHdb; \citealt{Boersma2015, Mattioda2020}), band-ratio analysis remains the conventional (if limited) standard in the \textit{James Webb} Space Telescope (JWST) era \citep{Maragkoudakis2022, Rigopoulou2024, Gregg2026}. 

\begin{figure}[htbp]
\centering
\includegraphics[width=8.8cm]{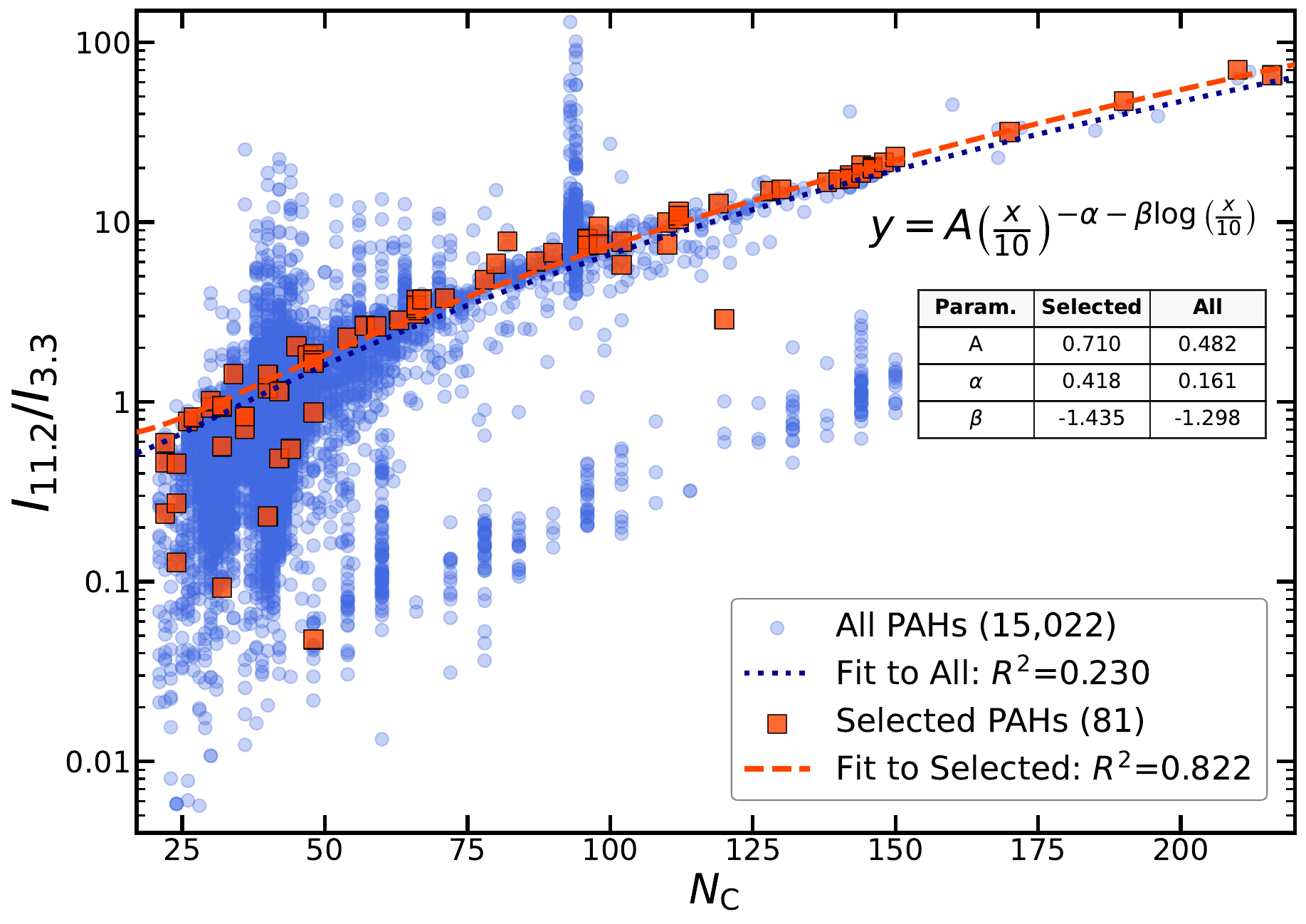}
\caption{Emission intensity ratio ($I_{11.2}/I_{3.3}$) vs. the number of carbon atoms ($N_{\mathrm{C}}$). Circles represent the full dataset of 15 022 neutral PAHs. Squares denote the subset of 81 species selected by \cite{Maragkoudakis2020}, characterized by $N_{\mathrm{C}} > 20$, the presence of solo C-H bonds, and the absence of heteroatoms. All spectra were re-computed using a 6 eV cascade model. The lines depict the fit where $R^{2}$ values highlight the fitting quality.
\label{F1}}
\end{figure}

However, Fig.~\ref{F1} demonstrates a critical shortcoming of the ratio-based paradigm: its perceived accuracy is often a byproduct of sample selection, rather than physical universality \citep{Maragkoudakis2020}. For instance, the $I_{11.2}/I_{3.3}$ size trend performs well for a specific subset of 81 molecules ($R^{2}=0.82$), yet it collapses and becomes unreliable ($R^{2}=0.23$) when applied to a comprehensive library of \num{15022} neutral PAHs. This discrepancy highlights a fundamental mismatch between classical diagnostic methods and modern observations. With JWST's unprecedented spectral resolution and sensitivity, reducing complex spectral profiles to a few discrete ratios is no longer a necessary simplification, but it comes at the cost of discarding valuable diagnostic information. To fully exploit the capabilities of current and future observatories, it is essential to transition from empirical ratios toward full-spectrum inference methods.

The advent of machine learning (ML) in astrophysics, coupled with expansive spectral libraries and JWST observations, presents a timely opportunity to move beyond traditional band-ratio diagnostics. In this work, we transition from discrete ratios to full-spectrum morphology, treating the complete IR spectrum as a high-dimensional fingerprint of PAH size and charge. Specifically, we trained an ML classifier on a large ensemble of spectra and evaluated its performance on observation-like mixtures of previously unseen species. This approach aims to provide more accurate and probabilistic constraints on molecular properties than conventional empirical methods. Instead of relying on degenerate band ratios, our method directly maps the full spectrum to radiation field hardness, PAH size, and charge, overcoming degeneracies, such as that seen between ionization and radiation fields \citep{Rigopoulou2021}.

\section{Methodology}

To streamline our analysis, we built a data-driven pipeline to infer PAH size and charged directly from their IR emission spectra. It integrates a large dataset of emission spectra with a ML classifier. While the model is trained on spectra of individual molecules, its performance is validated on complex, observation-like spectral mixtures drawn from an independent ``unseen'' molecular pool.

The dataset comprises \num{23653} unique PAH structures compiled from PAHdb (\num{10404} theoretical spectra, $6 \leq N_{\mathrm{C}} \leq 384$, charges -1, 0, +1, +2; \citealp{Ricca2026}) and first-principles DFT calculations (\num{13986} spectra, $8 \leq N_{\mathrm{C}} \leq 160$, charges -1, 0, +1; \citealp{He2026, Meng2023}). After de-duplication, the raw dataset spans \num{23653} PAH structures with harmonic IR spectra from 6.95 to \SI{3751}{\per\centi\meter}, including \num{15911} neutrals (\num{15022} with $N_{\mathrm{C}}>20$), \num{2047} anions, \num{2972} cations, and \num{2723} dications.

The molecules were categorized into a 12-class framework based on size and charge state. For size: small ($N_{\mathrm{C}} < 50$), medium ($50 \le N_{\mathrm{C}} \le 99$), and large ($N_{\mathrm{C}} \ge 100$). For charge state: anion (-1), neutral (0), cation (+1), and dication (+2).

These spectra represent ground-state absorption. To simulate astrophysically relevant conditions, we converted them to emission spectra using the thermal-cascade approximation within the AmesPAHdbIDLSuite tool \citep{Boersma2013, Boersma2014}, assuming a representative excitation energy of \SI{6}{eV}.

To maintain consistency with traditional theory, we focused on the 2.76--\SI{20}{\micro\meter} window (500--\SI{3620}{\centi\meter^{-1}}), which encompasses the characteristic PAH IR bands. Each discrete line spectrum was converted into a fixed-length feature vector by binning onto a common histogram grid with a fixed bin width and normalizing to unit area, so the model learns spectral shape rather than absolute intensity. Following sensitivity testing (see Appendix \ref{A1}), we used a bin width of \SI{20}{\centi\meter^{-1}} and excluded any features (bins) containing contributions from fewer than ten molecules. While this bin width optimizes the model's signal-to-noise ratio (S/N) and generalization potential, we acknowledge that this resolution may smooth out fine-grained spectral substructures potentially resolvable by JWST. This choice prioritizes robust morphological patterns over noise-sensitive narrow-band variations.

To ensure the model's astrophysical relevance, we partitioned the data by randomly selecting 20\% of the molecules from each of the twelve classes to form an ``unseen'' pool, while the remaining 80\% constituted the training set. We simulated realistic astronomical observations by constructing synthetic mixtures, averaging the spectra of molecules drawn at random from the unseen pool. This uniform weighting ensures the model remains independent of specific astrophysical priors. These mixtures were generated across varying population sizes, defined as $N_{\mathrm{mol}} \in \{1, 5, 10, 20, 50, 100, 200\}$. For each $N_{\mathrm{mol}}$ value, we produced 100 mixed spectra, resulting in 700 synthetic spectra per class. All mixtures derived from this unseen subset were reserved strictly for final model evaluation to ensure unbiased performance metrics.

We adopted a random forest (RF) classifier, an ensemble of decision trees that combines numerous threshold-based rules to produce robust predictions \citep{Breiman2001}. The model was trained on normalized spectral features using 500 trees with a maximum depth of 25, utilizing out-of-bag scoring for internal validation. To mitigate class imbalance, we implemented the synthetic minority oversampling technique (SMOTE) in conjunction with adjusted class weighting. The trained classifier was evaluated on synthetic mixtures derived from the ``unseen'' molecular pool. Performance was quantified using standard metrics: precision, recall, and $F_{1}$ score. To ensure reproducibility, the source code, datasets, and the trained model are openly accessible on Git repository: \href{https://github.com/zwAstroChem/AstroPAH-MLDiag}{AstroPAH-MLDiag}. Comprehensive hyperparameter configurations can be found in this code.

\section{Results and discussion}

\begin{table}[htbp] 
\caption{Performance metrics of PAH mixture classification across 12 size and charge categories.}
\label{T2}
\centering
\begin{tabular}{c S[table-format=5.0] ccc} 
\hline\hline
Category & {Training size} & Precision & Recall & $F_{1}$ score \\
\hline
Small  & 13626 & 0.985 & 0.983 & 0.985 \\
Medium & 4638  & 0.910 & 0.938 & 0.923 \\
Large  & 663    & 0.975 & 0.992 & 0.983 \\
\hline
$-1$   & 1639  & 0.970 & 0.983 & 0.977 \\
$0$    & 12730 & 0.893 & 0.973 & 0.930 \\
$+1$   & 2379  & 0.977 & 0.980 & 0.977 \\
$+2$   & 2179  & 0.987 & 0.947 & 0.963 \\
\hline
{Average} & {---} & {0.957} & {0.971} & {0.963} \\
\hline
\end{tabular}
\tablefoot{Results are grouped by molecular size ($N_{\mathrm{C}} < 50$, $50 \text{--} 99$, and $\ge 100$) and charge state ($-1$, $0$, $+1$, and $+2$).}
\end{table}

The trained RF model shows high-fidelity performance on the mixed PAH spectra, achieving a macro-averaged $F_{1}$ score of approximately 0.963 (Table \ref{T2}). This robust classification indicates that full-profile spectral features contain sufficient information to simultaneously constrain molecular size and ionization state, even within complex molecular mixtures.

The performance scales nonlinearly with molecular size. Overall, small PAHs ($N_{\mathrm{C}} < 50$) achieve the highest $F_{1}$ score (0.985), supported by a large training set of \num{13626} spectra. Large PAHs ($N_{\mathrm{C}} \geq 100$) perform nearly as well ($F_{1} = 0.983$) despite having only 663 training samples, which is the smallest among all size classes. This suggests that spectral features of large PAHs are sufficiently distinctive that the model requires relatively few examples to generalize accurately \citep{Bauschlicher2008, Ricca2012}. In contrast, medium-sized PAHs ($50 \le N_{\mathrm{C}} \le 99$) show a clear performance drop ($F_{1}$ = 0.923) even though they are reasonably well represented with \num{4638} training samples.

Charge classification performance is consistently high, with both cations ($+1$) and anions ($-1$) achieving $F_{1}$ scores of 0.977, and dications ($+2$) following at 0.963. The neutral class ($0$) exhibits the highest recall (0.973) but the lowest precision (0.893), indicating that while the model successfully identifies nearly all neutral species, it misassigns ionized or differently sized molecules to the neutral category.

\begin{figure}[htbp]
\centering
\includegraphics[width=8.8cm]{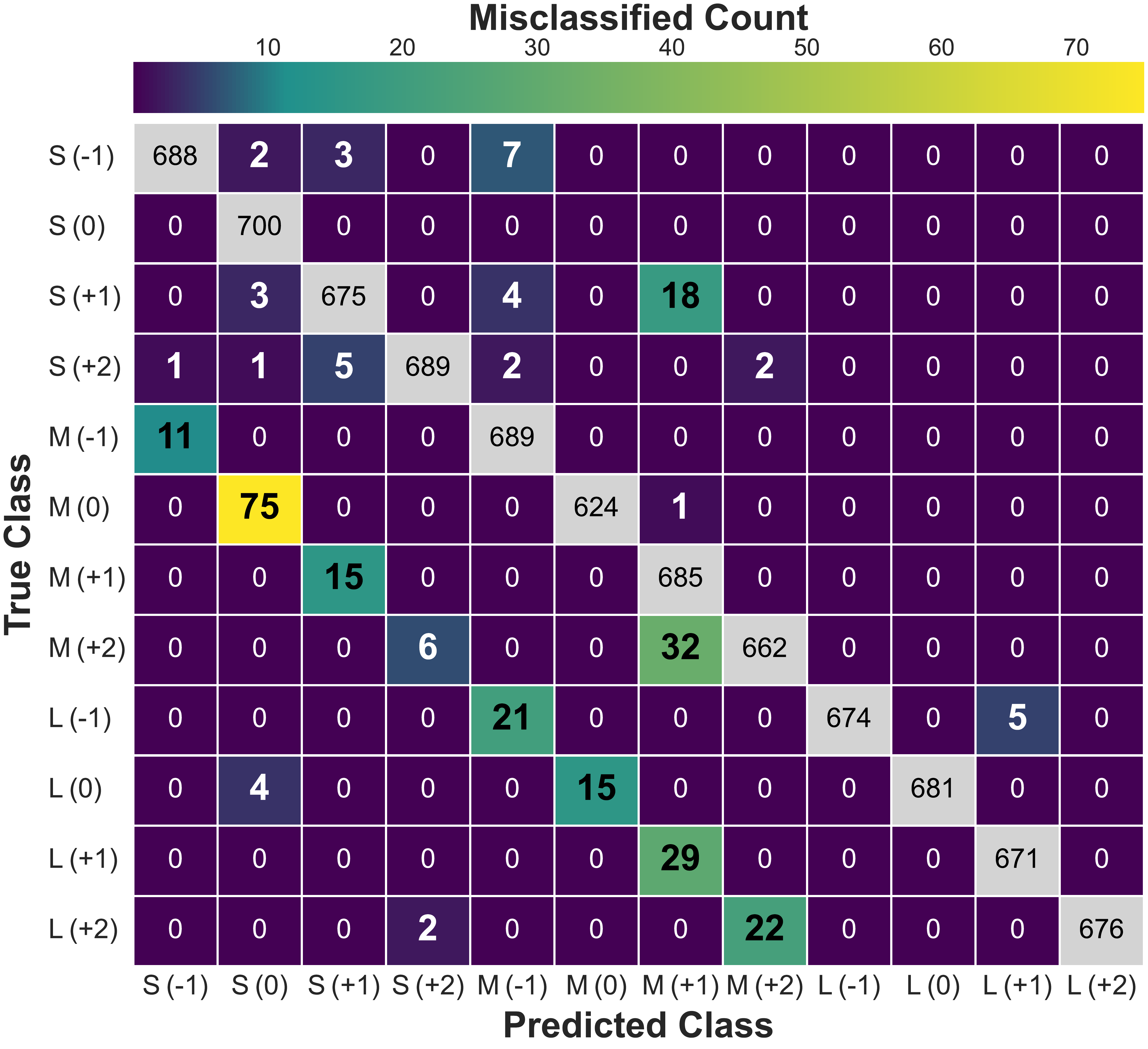}
\caption{Confusion matrix for the 12-class PAH mixture classification. Diagonal elements indicate correct classifications, while off-diagonal values reveal specific misclassification patterns among the 700 samples per class. Class indices are categorized by size: S (small), M (medium), and L (large), followed by the charge state in parentheses.}
\label{F2}
\end{figure}

To understand these performance dips, we can take a look at the confusion matrix (Fig. \ref{F2}), which reveals specific inter-class leakage. Within the medium-sized PAHs (M), misclassifications are dominated by neutral species [M\,(0)], which are frequently mistaken for small neutral [S\,(0)] PAHs, rather than ionized species. This confusion suggests that the spectral signatures of medium-sized neutrals begin to converge with those of smaller counterparts. As molecular size increases, rising vibrational mode density and spectral broadening are likely to diminish the distinctiveness between size classes \citep{Bauschlicher2009, Knight2021, Draine2007}. Given the high sampling rate for the medium category, this ambiguity appears to be a physical limitation of the spectral features rather than a lack of training data.

This trend extends across all charge states. Systematic misclassifications in Fig. \ref{F2} appear along the lines parallel to the diagonal, offset by exactly four classes, the interval corresponding to the same charge state in an adjacent size family. This pattern confirms that while the model readily distinguishes charge states, molecular size remains the more challenging attribute to constrain \citep{Draine2021}. Furthermore, charge-state residuals are confined to adjacent ionization levels; for example, 32 (out of 700) instances occurred where medium-sized  dications [M\,(+2)] were misidentified as cations [M\,(+1)].

To evaluate the influence of the training set size range on our results, we performed two sensitivity analyses. First, acknowledging that small PAHs ($N_{\mathrm{C}} < 20$) are susceptible to photodissociation in the diffuse interstellar medium \citep{Li2001}, we retrained the model with a minimum threshold of $N_{\mathrm{C}} = 20$; the resulting mean $F_{1}$ score remained stable at $0.947$, confirming that the inclusion of very small species does not bias performance for more resilient populations. Second, while computational costs currently cap our training data at $N_{\mathrm{C}} = 384$, we tested the model's so-called out-of-distribution generalizability by training on subsets with $N_{\text{C,max}} \leq 200$. The consistently high performance ($F_{1} > 0.95$) indicates that the model can effectively learn fundamental spectral motifs that remain characteristic in larger molecular structures ($N_{\mathrm{C}} \sim 1000$), as modeled in traditional frameworks \citep{Draine2007}.

\begin{figure}[htbp]
\centering
\includegraphics[width=8.8cm]{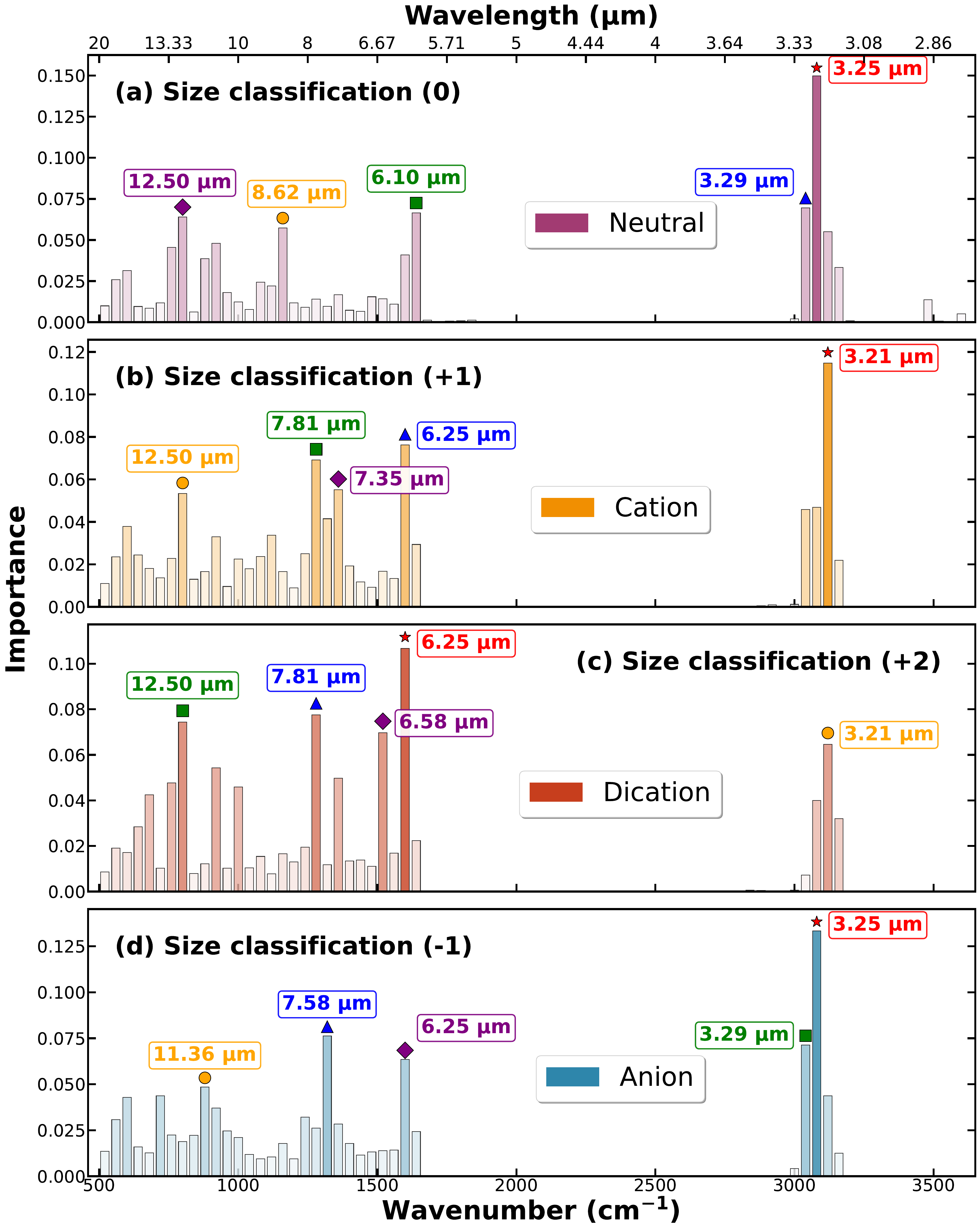}
\caption{Feature importance for size classification across different charge states: (a) neutral ($0$), (b) cation ($+1$), (c) dication ($+2$), and (d) anion ($-1$). Highlighted markers indicate the top five influential features.\label{F3}}
\end{figure}

The inherent interpretability of the RF model allows for an examination of its ``reasoning'' through a feature importance analysis. Using the Gini importance metric \citep{Breiman2001}, we identified the spectral regions most diagnostic of PAH properties. Figure~\ref{F3} shows that the size information is distributed across multiple spectral regions in a charge-dependent manner. For neutral PAHs (panel a), the model relies heavily on the 3.25 and \SI{3.29}{\micro\meter} bins, consistent with the classical proxy \citep{Lemmens2023}. The analysis also identifies the 6.1 (\ce{C-C} stretching), 8.6 (\ce{C-H} in-plane bending), and \SI{12.5}{\micro\meter} (\ce{C-H} out-of-plane (OOP) bending) bins as critical features for sizing the neutrals.

For cations (panel b), and especially dications (panel c), size diagnostic importance shifts toward the 6--\SI{8}{\micro\meter} \ce{C-C} stretching modes at 6.25 and \SI{7.81}{\micro\meter}. The ML model extracts size information within these complexes, likely tracing charge-specific modifications or edge geometries, but still supports caution against using the 6.2/7.7 ratio alone for size determination \citep{Peeters2002, Bauschlicher2008}. The analysis further reveals a significant departure from canonical diagnostics for anionic PAHs ( panel d), where size is primarily traced by the relative strengths of the \SI{3.3}{\micro\meter} complex and the \SI{7.58}{\micro\meter} \ce{C-C} stretch \citep{Allamandola1989, Schutte1993}. However, the high diagnostic importance of the \SI{7.58}{\micro\meter} \ce{C-C} stretch in anions should be interpreted with caution. Given the relatively small training pool for large anions (only 145 samples), the model’s reliance on this specific bin might partially reflect the limited structural diversity within this subset, rather than a universal physical law. 

Figure \ref{F3} suggests the heightened importance of the \SI{12.5}{\micro\meter} bin across multiple charge states (0, +1, and +2), identifying it as a robust complementary size tracer. This importance stems from the \ce{C-H} OOP bending modes of ``duo'' and ``trio'' hydrogen atoms, which are spatially stable across varying radiation fields \citep{Shannon2016}. The intensity of the 12.5--\SI{12.7}{\micro\meter} complex relative to the \SI{11.2}{\micro\meter} solo-H mode tracks the evolution of edge structure as PAHs grow in size \citep{Shannon2015, Maragkoudakis2023}. This aligns with longer-wavelength signatures proposed as alternatives for observations lacking \SI{3.3}{\micro\meter} coverage \citep{Draine2021, Berne2022}.

We further assessed the model’s applicability by testing it against JWST MIRI observations of NGC 7027 (see Appendix \ref{A2}). The model identified small-to-medium neutral PAHs as the dominant species, aligning with the observed spectral profile. However, the moderate confidence levels that we obtained in this work point to an important limitation: astronomical observations represent an integrated line-of-sight view of mixed PAH populations. Although our current model is designed as a classifier that identifies a single dominant species or state, real spectra are inherently mixtures. This mismatch naturally reduces the confidence of any single-label prediction. Additionally, the model remains limited by its training on discrete energy levels rather than on continuous interstellar radiation fields \citep{Li2024_ApJ, Li2024_MNRAS}. As a preliminary test, this work also shows that future high-confidence diagnostics will  require integrating multi-instrument data (e.g., NIRSpec and MIRI) to capture the full range of diagnostic vibrational modes. Moreover, we evaluated the impact of excitation conditions on classification performance by applying our workflow to emission spectra at 3 and 9 eV. The classification accuracy remained consistent (see Appendix \ref{A3}), demonstrating a robustness across diverse astrophysical regimes. 

\section{Conclusions}

We present a ML framework designed to infer the size and charge of interstellar PAHs directly from full-spectrum IR morphology, achieving a macro $F_{1}$ score of 0.963 across 12 categories. Moving beyond ``black-box'' AI, our feature-importance analysis reveals that PAH size diagnostics are fundamentally charge-dependent. Specifically, while traditional empirical proxies such as $I_{11.2}/I_{3.3}$ show diminished reliability across large, diverse datasets, our model identifies the $3.21$--$3.29$ and $11$--\SI{14}{\micro\meter} spectral regions as the most informative feature clusters for size inference. Notably, the 12.5 $\mu$m feature emerges as a physically grounded, versatile tracer that remains robust across multiple charge states, effectively breaking the degeneracies that limit single-band-ratio diagnostics.

The model currently relies on synthetic mixtures as proxies for unknown astronomical ground truths. Further development of this work will prioritize addressing class imbalances, particularly the under-representation of large molecules, which currently limits the statistical confidence in their identified diagnostic footprints \citep{Kovacs2020, mai2025, Tang2026}. From a broader perspective, applications to JWST observations of NGC 7027 show that future efforts will require regression-based mixed-population modeling, continuous excitation frameworks, and multi-instrument data.

\section*{Data availability}

Source code and datasets are available on Git repository: \href{https://github.com/zwAstroChem/AstroPAH-MLDiag}{AstroPAH-MLDiag}.
 \begin{acknowledgements}
The authors acknowledge financial support from: National Natural Science Foundation of China (Grant No. 12463005), Guangxi Natural Science Foundation under (Grant No. 2026GXNSFHA00640301), and Guangxi Talent Programme (Highland of Innovation Talents).
\end{acknowledgements}

\bibliographystyle{aa}

\FloatBarrier 
\clearpage

\begin{appendix}
\nolinenumbers
\section{Sensitivity test for varying bin widths}
\label{A1}

The spectral preprocessing requires balancing resolution against model robustness. We tested bin widths of 8.4 (determined by the Knuth Bayesian rule), 12, 20, 30, and \SI{40}{\centi\meter^{-1}}, with results summarized in Table~\ref{TA1-1}. 

\begin{table}[h!]
\centering
\caption{Comparison of model performance (average $F_{1}$) across different bin widths (6~eV dataset).}
\label{TA1-1}
\begin{tabular}{cc}
\toprule
{Bin Width (\si{\centi\meter^{-1}})} & {Macro Average $F_{1}$} \\
\midrule
8.4  & 0.90 \\
12   & 0.93 \\
20   & 0.96 \\
30   & 0.90 \\
40   & 0.92 \\
\bottomrule
\end{tabular}
\end{table}

Although finer binning (e.g., 8.4 or \SI{12}{\centi\meter^{-1}}) provides higher nominal resolution, the resulting high-dimensional feature space (more bins) makes the classifier sensitive to non-diagnostic spectral noise and minor vibrational shifts. 

\section{Test on JWST MIRI observations}
\label{A2}

\begin{figure}[h]
\centering
\includegraphics[width=0.48\textwidth]{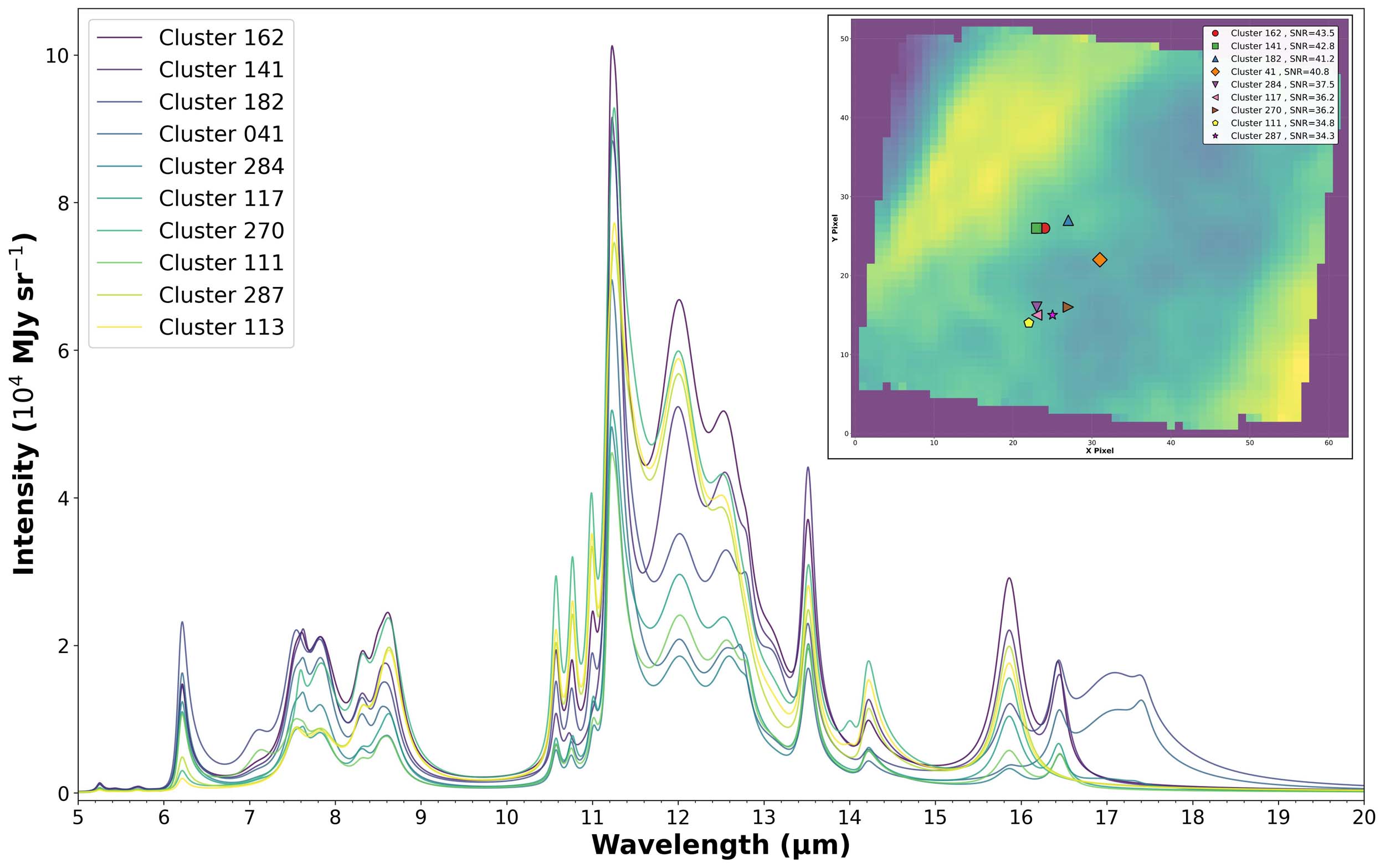}
\caption{JWST spectra of ten distinct PAH clusters in NGC 7027 over the 5–20 $\mu$m wavelength range. Inset: spatial distribution of the 10 PAH spectra in NGC 7027, where the colors of the background correspond to the mean flux.}
\label{FS1}
\end{figure}

We selected ten high signal-to-noise ratio (S/N $> 30.0$) spectra from the planetary nebula NGC 7027 (JWST Program 1523, PI: D. R. Law), as shown in the inset of Fig. \ref{FS1}. To isolate the PAH emission, we applied the differential extinction method of \citet{Donnan2024}, which extracts intrinsic spectral features. The resulting spectra are presented in Fig. \ref{FS1}.

We trained a separated RF classifier to predict the excitation energy across three classes: 3, 6, or \SI{9}{eV}. When evaluated on synthetic mixed spectra, the model achieved $F_{1}$ scores of about 0.989, 0.947, and 0.955 for the 3, 6, and \SI{9}{eV} classes, respectively. Using this classifier, we estimated the excitation environment for the observed spectra to be approximately 3.0 eV.

We then applied our 12-class ML framework (re-trained with 3-eV dataset) to determine PAH size and charge states. The model identifies the carriers as small-to-medium neutral PAHs, which is qualitatively consistent with the observed spectral profiles (e.g., the prominent 11.3 $\mu$m feature relative to the 6.2, 7.7, and 8.6 $\mu$m bands, relatively well-resolved discrete features in 11--15 $\mu$m region, and weak feature at 17 $\mu$m). However, we noted that the classification confidence remains low. For the ten selected PAH clusters in NGC 7027, the model predicts eight clusters (IDs: 111, 141, 117, 113, 270, 284, 162, and 287) as small neutral PAHs (Small (0)), with confidence scores of 0.696, 0.654, 0.650, 0.598, 0.574, 0.572, 0.568, and 0.564, respectively; along with two clusters (IDs: 041 and 182) as medium neutral PAHs (Medium (0)), with confidence scores of about 0.578.

\section{Impact of excitation energy}
\label{A3}
\begin{figure}[htbp]
\centering
\includegraphics[width=8cm]{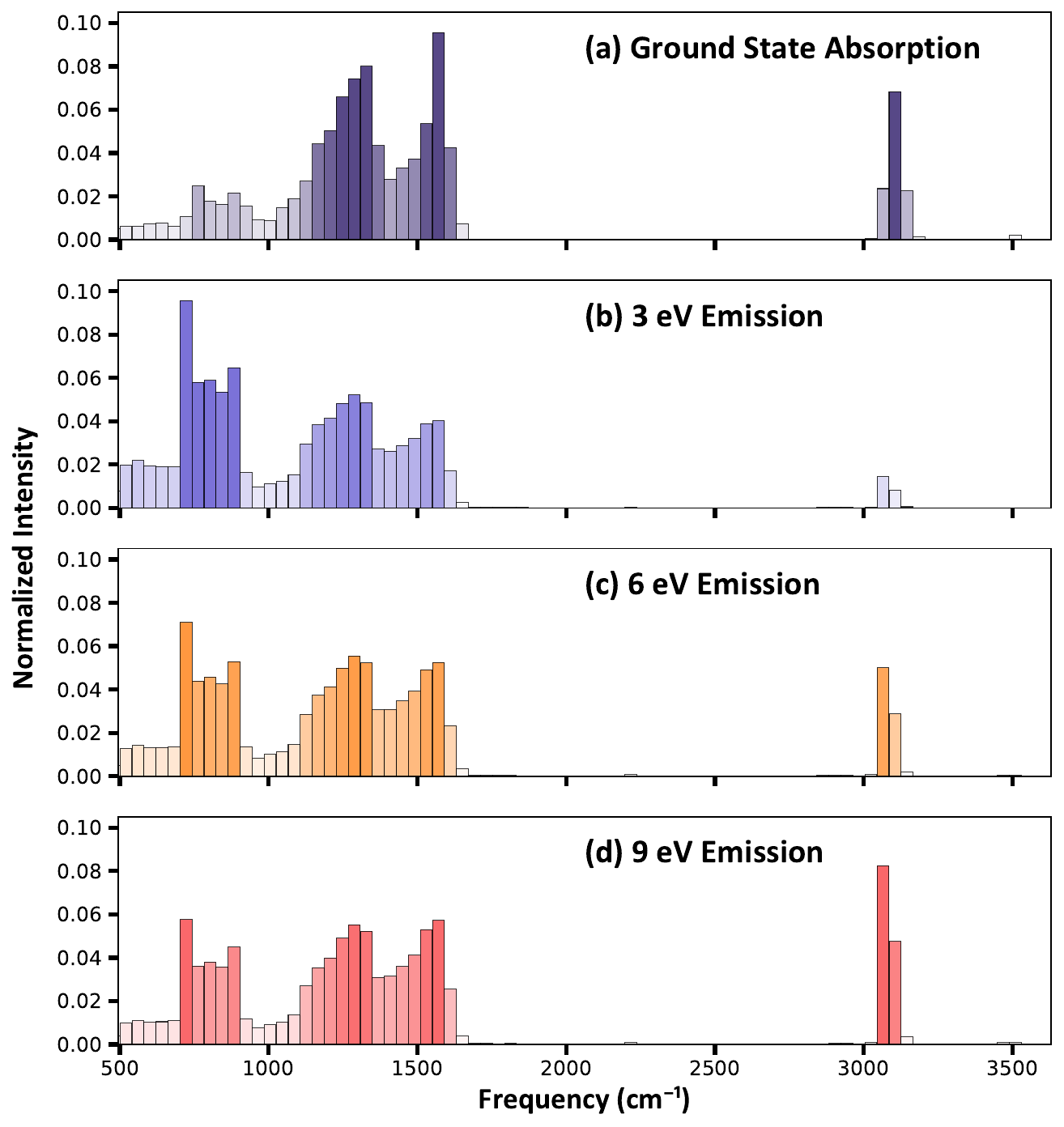}
\caption{Normalized summed emission spectra of all \num{23653} PAHs at excitation energies of 3 (b), 6 (c), and \SI{9}{eV} (d), shown alongside their ground-state absorption spectrum (a). Each spectrum was normalized with the total area scaled to unity. 
}
\label{FS2}
\end{figure}

We tested model performance across three excitation energies (3, 6, and \SI{9}{eV}) by simulating PAH emission spectra from absorption data via a thermal cascade model (Fig. \ref{FS2}).

As shown in Table~\ref{TA3-1}, classification performance peaks at \unit{6}{eV}. Performance remains high but shows a slight decrease at both \unit{3}{eV} and \unit{9}{eV} which yield Macro $F_{1}$ scores of 0.95 and 0.93.

\begin{table}[htbp]
\centering
\caption{Model performance metrics across different emission excitation energies.}
\label{TA3-1}
\begin{tabular}{lccc}
\toprule
{Spectral Type} & {Precision} & {Recall} & {$F_{1}$ score} \\ \midrule
\SI{3}{eV}         & 0.95                     & 0.95                  & {0.95}                    \\
\SI{6}{eV}          & 0.96            & 0.97         & {0.96}           \\
\SI{9}{eV}          & 0.94                     & 0.93                  & {0.93}                   \\ \bottomrule
\end{tabular}
\end{table}

\end{appendix}

\end{document}